\documentstyle[12pt]{article}
\def\ii{\'{\char'20}}

\begin{document}

\newcommand{\ep}{\epsilon}
\newcommand{\fr}{\frac}
\newcommand{\reals}{\mbox{${\rm I\!R }$}}
\newcommand{\nats}{\mbox{${\rm I\!N }$}}
\newcommand{\intgs}{\mbox{${\rm Z\!\!Z }$}}
\newcommand{\suml}{\sum_{l=0}^{\infty}}
\newcommand{\sknu}{\sum_{k=0}^{\infty}}
\newcommand{\cam}{{\cal M}}
\newcommand{\caz}{{\cal Z}}
\newcommand{\cao}{{\cal O}}
\newcommand{\caf}{{\cal F}}
\newcommand{\ing}{\int\limits_{\gamma}}
\newcommand{\imr}{\int\limits_{mR}^{\infty}dk\,\,}
\newcommand{\imrn}{\int\limits_{mR/\nu}^{\infty}dk\,\,}
\newcommand{\cah}{{\cal H}}
\newcommand{\nn}{\nonumber}
\newcommand{\komplex}{\mbox{${\rm I\!\!\!C }$}}
\newcommand{\spis}{\frac{\sin (\pi s)}{\pi}}
\newcommand{\pis}{\sin (\pi s)}
\newcommand{\snu}{\sum_{\nu =1/2,3/2,...}^{\infty}}
\newcommand{\numr}{\left(\frac{\nu}{mR}\right)^2}
\newcommand{\numu}{\left(\frac{\nu}{\mu}\right)^2}
\newcommand{\mzs}{m^{-2s}}
\newcommand{\xnur}{\int_{\frac{mR}{\nu}}^{\infty}\left(\left(\frac{x\nu}
R\right)^2-m^2\right)^{-s}}
\newcommand{\tpar}{\frac{\partial}{\partial \tau}}
\newcommand{\tzpar}{\frac{\partial^2}{\partial \tau^2}}

\newcommand{\abl}{\partial}
\newcommand{\kkk}{\frac{(-1)^k}{k!}}
\newcommand{\g}{\Gamma\left(}
\def\bce{\begin{center}}
\def\ece{\end{center}}
\def\beq{\begin{eqnarray}}
\def\eeq{\end{eqnarray}}
\def\brr{\begin{array}}
\def\err{\end{array}}
\def\ben{\begin{enumerate}}
\def\een{\end{enumerate}}
\def\bei{\begin{itemize}}
\def\eei{\end{itemize}}
\def\ul{\underline}
\def\ni{\noindent}
\def\bs{\bigskip}
\def\ms{\medskip}
\def\dsp{\displaystyle}
\def\wt{\widetilde}
\def\wh{\widehat}
\def\rl{$\Real$}
\def\tu{\bigtriangleup}
\def\td{\bigtriangledown}

\begin{titlepage}

\title{\begin{flushright}
{\normalsize UB-ECM-PF 95/1 }
\end{flushright}
\vspace{2cm}
{\Large \bf Heat-kernel coefficients of the Laplace operator on the
$3$-dimensional ball}}
\author{M. Bordag\\
Universit{\"a}t Leipzig, Institut f{\"u}r Theoretische Physik,\\
Augustusplatz 10, 04109 Leipzig, Germany\\
K. Kirsten\thanks{Alexander von Humboldt foundation
fellow, E-mail address: klaus@zeta.ecm.ub.es}\\
Departament d'ECM, Facultat de F{\ii}sica
\\ Universitat de Barcelona, Av. Diagonal 647, 08028 Barcelona \\
Spain}
\date{January 10, 1995}

\maketitle

\begin{abstract}
We consider the heat-kernel expansion of the massive Laplace operator on
the three dimensional ball with Dirichlet boundary conditions. Using
this example, we illustrate a very effective scheme for the calculation
of an (in principle) arbitrary number of heat-kernel coefficients for
the case where the basis functions are known. New results for the
coefficients $B_{\frac 5 2},...,B_5$ are presented.
\end{abstract}
\end{titlepage}
It is important to know the explicit form for the coefficients in the
short-time expansion of the heat-kernel, $K(t)$, for some Laplacian-like
operator on a $d$-dimensional manifold $\cam$. In mathematics the
interest stems for example from connections between the heat-equation
and the Atiyah-Singer index theorem \cite{gilkey1}, whereas in physics
the interest in this expansion lies for example in the domain of quantum
field theory where it is commonly known as the (integrated) Schwinger-De
Witt proper-time expansion \cite{birrell}.

If the manifold $\cam$ has a boundary $\partial \cam$, the coefficients
$B_n$ in the short time expansion have volume and boundary parts
\cite{greiner}. Thus \beq
K(t) \sim (4\pi t)^{-\frac d 2}\sum_{k=0,1/2,1,...}^{\infty}
B_kt^k\label{neu1}
\eeq
with
\beq
B_k =\int_{\cam}dV\,\,b_n+\int_{\partial \cam}dS\,\,c_n.\label{neu2}
\eeq
For the volume part effective systematic schemes have been developed
(see for example
\cite{volume}). The calculation of $c_n$ is in general more difficult.
Only relatively recently the coefficient $c_2$ for Dirichlet and Neumann
boundary conditions have been found
\cite{discr,boundary}. When using the general formalism of
ref.~\cite{discr} for higher spin particles, Moss and Poletti
\cite{moss1} found a discrepancy
with the direct calculations of D'Eath and Esposito \cite{giam1} (see
also \cite{giam2}). The latter results have been confirmed in
\cite{sasha}
where a new systematic scheme for the calculation of $c_2$ has
been developed in the context of the Hartle-Hawking wave-function of
the universe for the case when the full set of basis functions is known
\cite{sasha}. Finally, very recently the discrepancy have been resolved
\cite{vasil} and now the results found using the general algorithm
\cite{moss2} are in agreement with the direct calculations
\cite{giam1,giam2,sasha}.

We will use a variant of the approach of ref.~\cite{sasha} in order to
show
that higher coefficients $c_n$ may be calculated very effectively for
the case that the basis functions are known. To illustrate the method
in this letter we will concentrate on the calculation of the
heat-kernel coefficients of the elliptic operator
\beq
(-\Delta +m^2) \phi _{n,l,m} = (\lambda ^2_{n,l,m} +m^2) \phi _{n,l,m}
\label{1}
\eeq
in the three dimensional ball defined by $B_R=\{\vec x \in \reals ^3,
|\vec x|\leq R\}$. We will treat explicitly Dirichlet boundary
conditions, $\phi_{n,l,m} (|\vec x|=R)=0$.
As will be clear afterwards, other boundary conditions and higher
dimensional balls may be treated in exactly the same way.

Starting point is the equation
\beq
J_{l+\frac 1 2} (\lambda _{n,l,m}R) =0\label{2}
\eeq
for the eigenvalues $\lambda_{n,l,m}$, which are $(2l+1)$-times
degenerated. We are especially interested in the calculation of the
heat-kernel coefficients defined in equation (\ref{neu1}).
Instead of calculating the heat-kernel coefficients itself, we will
concentrate on the zeta function of the operator, eq.~(\ref{1}), and
recover the heat-kernel coefficients using the relations \cite{voros}
\beq
Res\,\,\zeta (s) = \frac{B_{\frac m 2 -s}}{(4\pi )^{\frac m 2} \Gamma
(s)}\label{3}
\eeq
for $s=\frac m 2,\frac{m-1}2,...,\frac 1 2;-\frac{2l+1}2,$ for $l\in
\nats_0$, and
\beq
\zeta (-p) = (-1)^p p! \frac{B_{\frac m 2 +p}}{(4\pi )^{\frac m 2}}
\label{4}
\eeq
for $p\in \nats _0$.

Using relation (\ref{2}) for the eigenvalues, one may write the zeta
function in the form (for a similar procedure see \cite{michael})
\beq
\zeta (s)& =&
\sum_{n=0}^{\infty}\suml \sum_{m=-l}^l \lambda _{n,l,m} ^{-2s}\nn\\
&=&\suml (2l+1) \ing\frac{dk}{2\pi i} (k^2+m^2)^{-s}\frac
{\partial}{\partial k}\ln J_{l+\frac 1 2}(kR),\label{5}
\eeq
where the contour $\gamma$ is counterclockwise enclosing all eigenvalues
which are known to be situated on the positive real axis. As it
stands, the representation (\ref{5}) is valid for $\Re s >\frac 3 2$.

Before considering in detail the $l$-summation, let us first construct
an analytic continuation of the $k$-integral in equation (\ref{5})
alone and let us define for that reason
\beq
\zeta _{\nu}(s)& =&
 \ing\frac{dk}{2\pi i} (k^2+m^2)^{-s}\frac
{\partial}{\partial k}\ln J_{\nu}(kR),\label{5a}
\eeq
with $\nu =l+1/2$.
Deforming the contour to the imaginary axis, the analytic
continuation,
\beq
\zeta _{\nu}(s) = \frac{\sin (\pi s)}{\pi} \imr\left(
\left(\frac k R
\right)^2-m^2\right)^{-s}\frac{\partial}{\partial k}\ln I_{l+\frac 1 2
}(k),\label{6}
\eeq
valid in the strip $\frac 1 2 < \Re s <1$, may be found. A similar
representation valid for $m=0$ has been given in \cite{sergio}.

In order to continue, the idea is to make use of
the uniform expansion of the
Bessel function  $I_{\nu} (k)$ for $\nu \to \infty$ as $z=k/\nu$ fixed
\cite{2}. Actually we use the expansion
\beq
\frac{\partial}{\partial k} \ln I_{\nu} ( k ) \sim \frac 1
{\sqrt{1-t^2} }
\sum_{n=0}^{\infty}\frac{d_n (t)}{\nu ^{n}}\label{7}
\eeq
with $t=\frac 1 {\sqrt{1+z^2}}$.
Here the functions $d_n (t)$ fulfill the recurrence
relation
\beq
d_n (t) =\frac 1 2 t (1-t^2) \left( t \frac{\partial}{\partial
t}-1\right) d_{n-1} (t) -\frac 1 2 \sum_{k=1} ^{n-1} d_k (t) d_{n-k}(t),
\label{11}
\eeq
starting with $d_0 (t) =1$.
In order to calculate up to $B_5$ one needs the first eleven
coefficients $d_n$, which can be easily calculated using the recurrence.

Adding and subtracting the leading terms of the asymptotic expansion
for $\nu\to\infty$,
eq.~(\ref{6}) may be split into two pieces,
\beq
\zeta _{\nu}(s) = N_{\nu}(s) +\sum_{i=1} ^{N+1} A_{\nu}^i (s),\label{8}
\eeq
with
\beq
N_{\nu}(s) =
\frac{\sin (\pi s)}{\pi}  \imrn\left( \left(\frac {k\nu} R
\right)^2-m^2\right)^{-s}\left\{\frac{\partial}{\partial k}\ln
I_{l+\frac 1 2
}(k\nu)-\frac 1 {\sqrt{1-t^2}}\sum _{n=0}^N \frac{d_n (t)}{\nu
^{n}}\right\} \label{9}
\eeq
and
\beq
A_{\nu}^i(s) =
\frac{\sin (\pi s)}{\nu^{i-1}\pi} \imrn\left( \left(\frac {k\nu} R
\right)^2-m^2\right)^{-s}\frac{\partial}{\partial k}d_{i-1}
(t).\label{10} \eeq
As it stands, the $A_{\nu}^i$, equation (\ref{10}), are well defined (at
least) in the strip $1/2 <\Re s<1$. However, the analytic continuation
in the parameter $s$ to the whole complex plane in terms of known
function may be provided.
To explain afterwards some details of the calculation let us give
explicitly only the $A_{\nu}^i $ of the two leading terms in the
asymptotics (\ref{7}),
\beq
A_{\nu}^1 (s)&=& \frac{m^{-2s}\sin (\pi s)}{2\pi ^{\frac 3 2}} Rm\nn\\
& &\times \qquad \Gamma \left(s-\frac 1 2\right) \Gamma (1-s)
{}~_2F_1\left(-\frac 1 2,s-\frac 1 2;\frac 1
2;-\left(\frac{\nu}{mR}\right)^2\right),\label{anu1}\\
A_{\nu}^2 (s) &=& -\frac{m^{-2s}\sin (\pi s)}{4\pi} \nn\\
& &\times \qquad \Gamma (s) \Gamma (1-s)
{}~_2F_1\left( 1,s;1;
-\left(\frac{\nu}{mR}\right)^2\right).\label{anu2}
\eeq
Similar expressions for higher $A_{\nu} ^i$ can be calculated,
e.g.~using standard integration packages. As mentioned,
we did explicitly calculate the coefficients up to $B_5$ and thus needed
$A_{\nu}^i$ for $i=1,...,11$.
Here $_2F_1(a,b,c;z)$ denotes the hypergeometric function \cite{1}.

The representation (\ref{8}) has the following very important
properties. First of all, by considering the asymptotics of the
integrand in equation (\ref{9}) for $k\to mR/\nu$ and $k\to \infty$, it
may be seen that \beq
N(s) =\suml (2l+1)N_{l+\frac 1 2} (s)\nn
\eeq
is analytic in the strip $1-N/2 < \Re s <1$. For that reason it gives no
contribution to the residue of $\zeta (s)$ in that strip.
Furthermore, for $s=-k$, $k=0,1,2,3$, we have $N(s) =0$ and thus it does
also not
contribute to the values of the zeta function at those points. Together
with eqs.~(\ref{3}), (\ref{4}), this yields, that the heat-kernel
coefficients are only determined by the terms $A_i (s)$ with
\beq
A_i (s) =\suml (2l+1)A_{l+\frac 1 2}^i (s).\label{17a}
\eeq
However, the $A_i (s)$ may be given in terms of Hurwitz
zeta functions and an explicit representation of
$A_i (s)$ showing the meromorphic structure in the whole complex
plane may be given.
The sum in (\ref{17a}) may be easily done by means of the
Mellin-Barnes type integral representation of the hypergeometric
function,
\beq
_2F_1(a,b;c;z) = \frac{\Gamma (c)}{\Gamma (a)\Gamma(b)}\frac 1 {2\pi i}
\int_{{\cal C}}dt\,\,\frac{\Gamma (a+t)\Gamma (b+t)\Gamma (-t)}{\Gamma
(c+t)} (-z)^t,\label{13}
\eeq
where the contour ${\cal C}$ is such that the poles of $\Gamma (a+t)$
and $\Gamma (b+t)$ lie to the left of it and the poles of $\Gamma
(-t)$ to the right \cite{1}.
Defining
\beq
h(a,b,c;n) =\suml \left(l+\frac 1 2\right) ^n
{_2F_1}\left(a,b;c;-\left(\frac{l+\frac 1
2}{\mu}\right)^2\right)\label{14} \eeq
and closing the contour ${\cal C}$ to the left,
we arrive at
\beq
h(a,b,c;n) &=& \sum_{k=0}^{\infty}\frac{(-1)^k}{k!}
\mu^{2k}\times\label{15}\\
& &\left(\mu^{2a}\frac{\Gamma (b-a-k)\Gamma (a+k)}{\Gamma (c-a-k)}
\zeta_H \left(-n+2a+2k;\frac 1 2\right)\right.\nn\\
& &\left.+\mu^{2b}\frac{\Gamma (a-b-k)\Gamma (b+k)}{\Gamma (c-b-k)}
\zeta_H \left(-n+2b+2k;\frac 1 2\right)\right),
\nn
\eeq
which may be used for all summations we need for the calculation of the
$A_i$`s.
For example we obtain
\beq
A_1&=& \frac{m^{-2s}\sin (\pi s)}{\pi ^{\frac 3 2}}Rm \Gamma
\left(s-\frac 1 2\right)\Gamma (1-s)  h\left( -\frac 1 2,s-\frac 1
2,\frac 1 2;1 \right)\nn\\
   &=& R^{2s}\frac{\sin (\pi s)}{2\pi ^{\frac 3 2}}\Gamma (1-s) \suml
 \frac{(-1)^l}{l!}(mR)^{2l}\frac{\Gamma\left(l+s-\frac 1 2\right)}{s+l}
\zeta_H \left(2l+2s-2;\frac 1 2\right),\nn\\
A_2 &=& -\frac{m^{-2s} \sin (\pi s)}{2\pi} \Gamma (s) \Gamma
(1-s) h(1,s,1;1)\nn\\
    &=& -R^{2s}\frac{\sin (\pi s)}{2\pi}\Gamma (1-s) \suml
 \frac{(-1)^l}{l!}(mR)^{2l}\Gamma\left(l+s\right)
\zeta_H \left(2l+2s-1;\frac 1 2\right),\nn
\eeq
and similar expressions for the other $A_i$'s, $i=1,...,11$. The
sums appearing in the $A_i(s)$ are convergent for $|mR|<1/2$. Using
this representation, equations (\ref{3}), (\ref{4}), together with
\beq
\zeta_H \left(1+\ep ;\frac 1 2\right) &=&\frac 1 {\ep }+\cao (\ep
^0),\nn\\
\Gamma (\ep -n) &=& \frac 1 {\ep} \frac {(-1)^n}{n!} +\cao (\ep ^0),\nn
\eeq
the heat-kernel
coefficients may be easily determined.
Summarizing, we find the following new results for the coefficients
$B_{\frac 5 2},...,B_5$,
\beq
B_{\frac 5 2}&=&\pi^{\frac 3 2}\left(
\frac{m^2} 6 - \frac 1{120R^2} - m^4R^2\right)\nn\\
B_3& =&\pi\left( -\frac{64}{9009R^3} + \frac{16 m^2}{315R} +
\frac{4m^4R}3-
           \frac{2m^6R^3} 9\right)\nn\\
B_{\frac 7 2} &=&\pi^{\frac 3 2}\left( -\frac{m^4}{12} -
\frac{47}{20160R^4} + \frac{m^2}{120R^2} + \frac{m^6R^2} 3\right)\nn\\
B_4& =&\pi\left(- \frac{202816}{72747675R^5}+
   \frac{64m^2}{9009R^3} -
  \frac{8m^4}{315R}\right.\nn\\
& &\left. - \frac{4m^6R} 9 +
  \frac{m^8R^3}{18}\right)\nn\\
B_{\frac 9 2}&=&\pi^{\frac 3 2}\left(
\frac{m^6}{36} -\frac{521}{443520R^6}
+\frac{47m^2}{20160R^4}\right.\nn\\ & &\left.-\frac{m^4}{240R^2}-
 \frac{m^8R^2}{12}\right)\nn\\
B_5&=& \pi \left(-\frac{25426048}{15058768725R^7} +
\frac{202816m^2}{72747675R^5} -
\frac{32m^4}{9009R^3}\right.\nn\\
& &\left.+\frac{8m^6}{945R} +\frac{m^8R} 9
  -\frac{m^{10}R^3}{90}\right)\nn
\eeq
For $m=0$ the coefficient $B_{5/2}$ agrees with the one found by Kennedy
\cite{gerard}.

As mentioned, other boundary conditions, higher dimensional balls, and
even higher
coefficients may be found in exactly the same way without any additional
complication. This and more details of the calculation will be given in
a separate publication.
\vspace{5mm}

\ni{\large \bf Acknowledgments}

It is a pleasure to thank S.~Leseduarte, E.~Elizalde, P. Gilkey, S.
Dowker and G.~Esposito for interesting discussions and helpful comments.
K.K. thanks the Department ECM of the University of Barcelona for their
warm hospitality. Furthermore, K.K. acknowledges financial support from
the Alexander von Humboldt Foundation (Germany).

\end{document}